\documentclass[prd, preprint, 12pt]{revtex4-1}

\usepackage{amsmath}
\usepackage{amssymb}
\usepackage{setspace}
\usepackage{graphicx}
\usepackage{natbib}
\usepackage{float}

\begin{document}
 
 %

\begin{center}
 { \large {\bf Quantum Nonlocality, and the End of Classical Spacetime }}


\vskip 0.3 in

{\large{\bf Shreya Banerjee, Sayantani Bera and Tejinder P.  Singh}}

{\it Tata Institute of Fundamental Research,}
{\it Homi Bhabha Road, Mumbai 400005, India}\\
\bigskip
{\tt shreya.banerjee@tifr.res.in, sayantani.bera@tifr.res.in, tpsingh@tifr.res.in}\\

\end{center}

\bigskip
\bigskip

\centerline{\bf ABSTRACT}
\bigskip
\noindent Quantum non-local correlations and the acausal, spooky action at a distance suggest a discord between quantum theory and special relativity. We propose a resolution for this discord by first observing that there is a problem of time in quantum theory. There should exist a reformulation of quantum theory which does not refer to classical time. Such a reformulation is obtained by suggesting that space-time is fundamentally non-commutative. Quantum theory without classical time is the equilibrium statistical thermodynamics of the underlying non-commutative relativity. Stochastic fluctuations about equilibrium give rise to the classical limit and ordinary space-time geometry. However, measurement on an entangled state can be correctly described only in the underlying non-commutative space-time, where there is no causality violation, nor a spooky action at a distance.
\noindent 

\vskip 1 in

\centerline{March 20, 2016}

\bigskip

\centerline{Essay written for the Gravity Research Foundation 2016 Awards for Essays on Gravitation}
\bigskip

\centerline {{\bf Corresponding Author:} Tejinder P. Singh}
\bigskip
\centerline{\it This essay received an honorable mention in the Gravity Research Foundation 2016 Essay Contest}

\newpage

\setstretch{1.4}

\noindent{\it ``It may be that a real synthesis of quantum and relativity theories requires not }
 \noindent{\it just technical developments but radical conceptual renewal''.}
 \qquad\qquad \qquad\qquad\qquad {- J. S. Bell (1986)}
\bigskip

\noindent Measurements on entangled quantum states demonstrate non-local correlations and suggest the existence of an acausal action at a distance across space-like separated regions. This is confirmed by ever more precise loophole free tests of violation of Bell's inequalities by quantum systems \cite{loophole2015,loophole22015}. Even though such correlations cannot be used for superluminal signaling \cite{Ghirardi:80}, the acausal nature of the influence suggests a conflict with special relativity and Lorentz covariance \cite{Bell:86}. Numerous investigations over decades have not provided a satisfactory resolution of the problem. Further support for the conflict comes from a remarkable experiment showing that even if the influence was assumed to travel causally in a hypothetical privileged frame of reference, its speed would have be  at least four orders of magnitude greater than the speed of light \cite{spooky2008}!

A measurement causes the wave function to collapse instantaneously all over space. A dynamical theory such as Continuous Spontaneous Localization [CSL] explains collapse as a spontaneous physical process mediated by a stochastic nonlinear modification of the Schr\"odinger equation \cite{Ghirardi2:90}. Nonetheless, CSL is a nonrelativistic theory, and a satisfactory relativistic extension of CSL has not yet been found. We take the above-mentioned facts  as evidence that non-local quantum correlations and collapse of the wave function during a quantum measurement are not compatible with special relativity. In this essay we demonstrate how the resolution of the so-called problem of time in quantum theory enforces a non-commutative structure on space-time. We then show that  this non-commutative structure naturally explains entanglement and instantaneous wave function collapse across space-like separations. The same process appears unphysical and violates causality, when seen from the [incorrect] vantage point of Minkowski spacetime in special relativity. 


The time that appears in quantum theory is part of a classical space-time, whose geometry is determined by classical matter fields in accordance with the laws of the general theory of relativity. It is a consequence of the Einstein hole argument that in the absence of these classical fields, the manifold structure of space-time does not exist \cite{Carlip2001}. Thus, in its need for time, quantum theory has to depend on its own classical limit, and that is unsatisfactory from a fundamental viewpoint. There ought to exist an equivalent reformulation of quantum theory which does not refer to an external classical time \cite{Singh:2006}.  We have proposed that such a reformulation can be arrived at by generalizing the theory known as Trace Dynamics.


The goal of Trace Dynamcs [TD] is to derive quantum theory from a deeper underlying theory \cite{Adler:04}. This is more satisfactory, as compared to arriving at quantum theory by imposing ad hoc canonical commutation relations on a previously known classical mechanics. TD is the classical dynamics of $N\times N$ matrices $q_r$ whose elements are either odd grade [fermionic sector F] or even grade [bosonic sector B] elements of Grassmann numbers. The Lagrangian of the theory is defined as the trace of a polynomial function of the matrices and their time derivatives, and then Lagrangian and Hamiltonian dynamics can be developed in the usual way. The configuration variables $q_r$ and their conjugate momenta $p_r$ all obey arbitrary commutation relations amongst each other.  Nonetheless, as a consequence of a global unitary invariance there occurs in the theory a remarkable conserved charge, known as the Adler-Millard charge
\begin{equation}
\tilde{C} = \sum_B [q_r,p_r] -\sum_F \{q_r,p_r\} 
\end{equation}
whose existence is central to the subsequent development \cite{Adler-Millard:1996}.


Next, assuming that one is not examining the dynamics at this level of precision, one develops an equilibrium statistical thermodynamics of the classical dynamics described by TD. The equipartition of the Adler-Millard charge implies certain Ward identities, which lead to the important result that thermal averages of canonical variables obey quantum dynamics and quantum commutation relations \cite{Adler:04}. In particular, the emergent $q$ operators commute with each other, and so do the $p$ operators. Furthermore, if one considers the inevitable statistical fluctuations of the Adler-Millard charge about equilibrium, this leads to a CSL type modification of the nonrelativistic Schr\"odinger equation. The said modification, negligible for microscopic systems but significant for large objects, solves the quantum measurement problem and leads to emergent classical behavior in macroscopic systems \cite{RMP:2012}. The fluctuations of the conserved charge about its equilibrium value carry information about the arbitrary commutation relations amongst the configuration and momentum variables in the underlying TD.


In order to arrive at a formulation of quantum theory without classical time, we first generalized Trace Dynamics so as to make space-time coordinates also into operators \cite{Lochan-Singh:2011}. Associated with every degree of freedom there are coordinate operators  $(\hat{t}, \hat{\bf x})$ with arbitrary commutation relations amongst them. These define a Lorentz invariant line-element $d\hat{s}^2$, and the important notion of Trace time $s$ as follows:
\begin{equation}
ds^2 = Tr d\hat{s}^2 \equiv Tr[d\hat{t}^2 - d\hat{x}^2 - d\hat{y}^2 - d\hat{z}^2]
\label{nsr}
\end{equation}
A Poincar\'e invariant dynamics can be constructed, in analogy with special relativity, and in analogy with TD, but with evolution now defined with respect to trace time $s$. The theory continues to admit a conserved Adler-Millard charge, and the degrees of freedom now involve bosonic and fermionic components of space-time operators as well. {\it Because the space-time operators have arbitrary commutation relations, there is no point structure or light-cone structure, nor a notion of causality, despite the line-element being Lorentz invariant} \cite{Lochan-Singh:2011}.

Given this generalized TD, we construct its equilibrium statistical thermodynamics, as before. The equipartition of the Adler-Millard charge leads to the emergence of a generalized quantum dynamics [GQD] in which evolution is with respect to the trace time $s$, and the thermally averaged space-time operators $(\hat{t}, \hat{\bf x})$ are now a subset of the configuration variables of the system \cite{Lochan:2012}. It is important to note that these averaged operators commute with each other. This is the sought after reformulation of quantum theory which does not refer to classical time. In the non-relativistic limit one recovers the generalized 
Schr\"odinger equation
\begin{equation}
i\hbar \frac{d\Psi(s)}{ds} = H\Psi (s)
\label{gqd}
\end{equation}


Before we demonstrate the equivalence of the reformulation with standard quantum theory, we must explain how the classical Universe, with its classical matter fields and ordinary space-time, emerges from the GQD in the macroscopic approximation. Like in TD, one next allows for inclusion of stochastic fluctuations of the Adler-Millard charge, in the Ward identity. This again results in a non-linear stochastic Schr\"odinger equation, but now with important additional consequences. Consider the situation where matter begins to form macroscopic clumps (for instance in the very early universe, soon after the Big Bang). The stochastic fluctuations become increasingly significant as the number of degrees of freedom in the clumped system is increased. These fluctuations induce macroscopic objects to be localized, but now not only in space, but also in time! This means that the time operator associated with every object becomes classical (a $c$-number times a unit matrix) \cite{Singh:2012}.

The localization of macroscopic objects is accompanied by the emergence of a classical space-time, in accordance with the Einstein hole argument.  If, and only if, the Universe is dominated by macroscopic objects, as the Universe today is, can one also talk of the existence of a classical space-time. When this happens, the proper time $s$ may be identified with classical proper time. Once the Universe reaches this classical state, it sustains itself therein, by virtue of the continuous action of stochastic fluctuations on macroscopic objects, thereby achieving also the existence of a classical space-time geometry \cite{Singh:2012}. Because the underlying generalized TD is Lorentz invariant, the emergent classical  space-time is locally Lorentz invariant too. However there is a key difference: unlike in the underlying theory, now light-cone structure and causality emerge, because the space-time coordinates are now $c$-numbers.


[We note that in conventional studies of the very early Universe, gravity is assumed to become classical at
the Planck scale, while matter fields are assumed to become classical later. Clearly, such a scenario is feasible only if the semiclassical gravity approximation is valid, e.g. $G_{\mu\nu}=\kappa\langle \Psi|T_{\mu\nu}|\Psi\rangle$, which typically requires matter to be in a highly coherent, nearly classical state. Something like this is implicitly assumed, for example, during the inflationary epoch, with the inflaton field being composed of a dominant classical part, and a sub-dominant quantum perturbation].

Independent of this pre-existing classical background, a microscopic system in the laboratory is fundamentally described, on its own non-commutative space-time (\ref{nsr}), by the associated generalized TD. Upon coarse-graining, this leads to the system's GQD (\ref{gqd}) with its trace time. Under the assumption that stochastic fluctuations can be ignored, this GQD has commuting $\hat t$ and $\hat {\bf x}$ operators. These, by virtue of their commutativity, can be mapped to the $c$-number $t$ and ${\bf x}$ of the pre-existing classical universe, and trace time can be mapped to ordinary proper time. This is a map to ordinary special relativity, and hence one recovers standard relativistic quantum mechanics, and its non-relativistic limit. This shows how standard quantum theory is recovered from the reformulation which does not have classical time \cite{Singh:2012}.


We now have at hand all the ammunition needed to attack the spooky action at a distance, when a measurement is made on an entangled quantum state over space-like separations. Prior to the measurement, the stochastic fluctuations of the Adler-Millard charge can be neglected for the quantum system, and as we saw above, its GQD can be mapped to ordinary quantum theory. However, when the measurement is made, the collapse inducing stochastic fluctuations in the space-time operators $\hat{t}, \hat{\bf x}$ associated with the quantum system come into play. These operators now carry information about the arbitrary commutation relations of the underlying TD and no longer commute with each other. Hence they cannot be mapped to the ordinary space-time of special relativity.   Simultaneity can only be defined with respect to the trace time $s$, and there is no special relativistic theory of collapse. Collapse and the so-called non-local quantum correlation  truly takes place only in the non-commutative space-time (\ref{nsr}), which is devoid of point structure, devoid of light-cone structure, and devoid of the notion of distance. Hence one can only say that collapse takes place at a particular trace time, which is Lorentz invariant, and it is not meaningful to talk of an influence that has travelled, nor can one call the correlation non-local. 

If one tries to view and describe the measurement on the entangled quantum state from the vantage point of the Minkowski space-time of special relativity, the process naturally appears acausal and non-local. However, such a description is invalid, because there is no map from the fluctuating and noncommuting 
${\hat t}, \hat{\bf x}$ to the commuting $t$ and ${\bf x}$ of ordinary special relativity. There is no such map in the non-relativistic case either. However, in the non-relativistic case,  since there is an absolute time, it is possible to  model the fluctuations as a stochastic field on a given space-time background, like in CSL, and collapse is instantaneous in this absolute time, but does not violate causality.


We  conclude that the problem of time in quantum theory is intimately connected with the vexing issue of quantum non-locality and acausality in entangled states. Addressing the former compels us to revise our notions of space-time structure, which in turn provides a resolution for the latter. And it compels us to think  about quantum gravity in a new way.


\bigskip

\centerline{\bf REFERENCES}

\bibliography{biblioqmtstorsion}

\end{document}